\documentclass[twocolumn,prl,floatfix, superscriptaddress]{revtex4}
\usepackage{mathrsfs,graphicx,amsmath,amssymb,amsbsy}

\begin{document}

\title{Breached superfluidity via p-wave coupling}

\author{E. Gubankova}
\affiliation{Center for Theoretical Physics, Department of
Physics, MIT, Cambridge, Massachusetts 02139}

\author{E.G. Mishchenko}

\affiliation{Department of Physics, University of Utah, Salt Lake City, UT 84112}
\affiliation{Lyman Laboratory, Department of Physics, Harvard
University, MA 02138}

\author{F. Wilczek}
\affiliation{Center for Theoretical Physics, Department of
Physics, MIT, Cambridge, Massachusetts 02139}

\leftline{MIT-CTP \# 3528}
\begin{abstract}
Anisotropic pairing between fermion species with different fermi momenta opens
two-dimensional areas of gapless excitations, thus producing a spatially homogeneous state
with coexisting superfluid and normal fluids.
This breached pairing state
is stable and robust for arbitrarily small mismatch and weak p-wave
coupling.
\end{abstract}
\maketitle

{\it Introduction.} Recently there has been considerable interest in the possibility of new forms of 
superfluidity that could arise when one has attractive
interactions between
species with different fermi surfaces. This is stimulated by experimental developments 
in cold atom systems \cite{coldatoms} and by
considerations in high-density QCD  \cite{ARW}.  
Possible coexistence of superfluidity with gapless excitations is an especially
important qualitative issue.
Spatially homogeneous superfluid states that coexist with gapless modes at isolated points or lines in momentum space arise
in a straightforward way when
BCS theory is generalized to higher partial waves \cite{Mineev}.
Gapless states also are well known to occur in the presence
of magnetic impurities
\cite{Abrikosov} and, theoretically, in states with spontaneous breaking of translation symmetry \cite{LOFF},
where the gapless states span a two-dimensional fermi surface. Strong coupling between different bands also may lead
to zeros in quasiparticle excitations and gapless states \cite{Volovik}.
For spherically symmetric
(s-wave) interactions a spherically symmetric {\it ansatz\/} of this type naturally suggests itself when
one attempts to pair fermions of two different species with distinct fermi surfaces, and a pairing solution
can be found \cite{Sarma,LW,GLW,ABR,SH}.
The stability of the resulting state against phase separation \cite{Bedaque}
or appearance of a tachyon in the gauge field
(negative squared Meissner mass) \cite{WuYip}
is delicate, however \cite{FGLW,AKR}.
It appears to require some combination of
unequal masses, momentum-dependent pairing interactions, and long-range neutrality constraints. Here we
demonstrate another possibility: direction-dependent interactions, 
specifically p-wave interactions. In this case, stability appears
to be quite robust. It seems quite reasonable, intuitively, that expanding an existing (lower-dimensional)
locus of zeros into a two-dimensional
zone should
be significantly easier than producing a sphere of gapless excitations ``from scratch''.  We shall show that it
even occurs for arbitrarily small coupling and small Fermi surface mismatch.

Interactions relevant to pairing can be dominated by p-wave (or higher) harmonics under several circumstances.  
If the s-wave interaction is repulsive, it will not be subject to the Cooper instability, and will not induce pairing.  
The Cooper instability can be regarded as an enhancement of the effective interaction for attractive channels as one integrates out 
high-energy modes near the Fermi surface. Thus the effective Hamiltonian will come to resemble the form we assume 
if the interspecies interactions are repulsive in s-wave but attractive in p-wave. Fermi statistics forbids diagonal (intraspecies)  
s-wave interactions; if the higher partial waves are repulsive, or weakly attractive, the model discussed here will apply.   
In the context of cold atom systems, tuning to an appropriate p-wave Feshbach resonance, 
as recently reported in  \cite{Zhang}, can insure interspecies p-wave dominance. 

A crucial difference between the model we consider and the
conventional p-wave superfluid system, $^{3}$He, lies in the distinguishability of 
the paired species. So although there are two components,
there is no approximate quasispin symmetry, and no analogue of the fully gapped
B phase \cite{BW}. In the absence of a magnetic field $^{3}$He has an approximate $SO(3)_L\times SO(3)_S\times U(1)$ 
symmetry under separate spatial, rotations, spin rotations, and number, which is spontaneously broken to the diagonal $SO(3)_{L+S}$ in the B phase.   
The residual symmetry enforces a gap of uniform magnitude in all directions in momentum space.   
The systems we consider have quite different symmetry and breaking patterns,
for example $U(1)_{L_z}\times U(1)_A\times U(1)_B \rightarrow U(1)_{L_z + A +B}$ for two spin-polarized species A, B in a magnetic field, 
or  $SO(3)_L\times U(1)_A\times U(1)_B \rightarrow U(1)_{L_z + A +B}$ if the magnetic field can be neglected.  
The reduced residual symmetry allows for interesting direction-dependent structure in momentum space.  
(In the A phases  $^{3}$He pairs effectively as two separate single-species systems, which again is quite different from our set-up.)

Experimental realizations of $p$-wave interaction in cold atom
systems have been reported recently in Ref.~\cite{TRJB}. Feshbach
resonance in $p$-wave occurs between $^{40}$K atoms in $f=9/2$,
$m_f=-7/2$ hyperfine states. This is in contrast to the $s$-wave
resonance, which occurs between non-identical $f=9/2$, $m_f=-9/2$
and $f=9/2$, $m_f=-7/2$ states \cite{LRTB}.  A promising
system for possible observation of the $p$-wave breached pairing
superconductivity is a mixture of $f=9/2$, $m_f=-9/2$ and $f=9/2$,
$m_f=-7/2$ atoms $^{40}$K tuned into the repulsive side of the
$s$-wave Feshbach resonance. Different densities (Fermi momenta)
of $m_f=-9/2$ and $m_f=-7/2$ particles can be prepared using different magnitudes
of an initial additional magnetic field, which is then removed.
Large atomic relaxation times ensure that the created (metastable)
states will exist long enough to allow formation of a
superfluid phase.

{\it Model:} Having in mind cold atoms in a magnetic trap with
atomic spins fully polarized by a magnetic field,
we consider a model system with the two species of
fermions having the same Fermi velocity $v_F$, but  different
Fermi momenta $p_F\pm I/v_F$. The effective Hamiltonian is
\begin{equation}
\label{ham} H=\sum_{\bf p} [ \epsilon^A_{ p} a_{\bf p}^\dagger
a_{\bf p}+ \epsilon^B_{p} b_{-\bf p}^\dagger b_{\bf -p} -
\Delta_{\bf p}^{*} a_{\bf p}^\dagger b^\dagger _{-\bf p} -
\Delta_{\bf p} b_{-\bf p} a_{\bf p}]
\end{equation}
with $\epsilon^A_{p}=\xi_p +I,~\epsilon^B_{p}=\xi_p -I$,
$\xi_p=v(p-p_F)$, $\Delta_{\bf p}=\sum_{\bf k} V_{\bf p-k}\langle
a_{\bf k}^\dagger b^\dagger _{-\bf k}\rangle$. Here the attractive inter-species
interaction is $-V_{\bf p-k}$ within the ``Debye'' energy $2\omega_D$
around the Fermi surface ($\omega_D\gg I$), and the intra-species
interaction is assumed to be either repulsive or negligibly small.
Excitations of the Hamiltonian
(\ref{ham}), $E_{\bf p}^{\pm}=\pm \sqrt{\xi^2_p+\Delta_{\bf p}^2}
+I$, are gapless provided that there are areas on the Fermi
surface where $I>\Delta_{\bf p}$. The gap equation at zero
temperature,
\begin{equation}
\Delta_{\bf p}=\frac{1}{2}\sum_{\bf k}V_{\bf p-k}
\frac{\Delta_{\bf k}}{\sqrt{\xi^2_k+\Delta_{\bf k}^2}}~
\theta\left(\sqrt{\xi^2_k+\Delta_{\bf k}^2}-I\right),
\end{equation}
can be simplified by taking the integral over $d\xi_p$,
\begin{eqnarray}
\Delta_{\bf n} &=&\nu \int \frac{d \o_{\bf n'} }{4\pi}
V({\bf n,n'}) \Delta_{\bf n'} \left(\ln{\frac{1}{|\Delta_{\bf
n'}|}} \right. \nonumber\\
&+& \left.\Theta (I-|\Delta_{\bf n'}|)\ln{\frac{|\Delta_{\bf
n'}|}{I+\sqrt{I^2-\Delta^2_{\bf n'}}} } \right).
\label{gapeq}
\end{eqnarray}
where $\nu=p_F^2/(2\pi^2 v_F)$ is the density of states. In the last expression it is
assumed that $I$ and $\Delta_{\bf n}$ are dimensionless, and
scaled with the ``Debye'' energy: $I \to 2\omega_D I$,
$\Delta_{\bf n} \to 2\omega_D \Delta_{\bf n}$. In deriving
Eq.~(\ref{gapeq}) we neglect dependence of $V_{\bf p-k}$ on the
absolute values of ${\bf p}$ and ${\bf k}$; this is valid for
$\omega_D \ll E_F$.  At weak coupling we may linearize in the partial wave expansion,
$V({\bf n,n'})=\sum_{l,m} V_lY_{lm}({\bf n})Y^{\star}_{lm}({\bf n'})$.
Assuming p-wave dominance,
we parameterize
$V({\bf n,n'})=g ({\bf n}\cdot {\bf n'})$ with $g>0$.
P-wave gap parameters can arise in the forms $Y_{10}$ and $Y_{1\pm 1}$,
describing polar and planar phases respectively. 

{\it Polar phase:} $\Delta_{\bf n} \sim Y_{10}({\bf n})$.
We look for a solution in the form $\Delta_{\bf n}=\Delta
\cos({\bf n,z})$ where ${\bf z}$ is a fixed but arbitrary direction
(rotational symmetry is broken). 
The gap
equation becomes,
\begin{eqnarray}
&-&\frac{1}{\nu g} = \int\limits_0^{\pi/2} d\theta
\sin{\theta}\cos^2{\theta} \ln\left(\Delta\cos{\theta}\right) \\
&+& \int\limits_{\theta^*}^{\pi/2} d\theta \sin{\theta}
\cos^2{\theta} \ln\left(\frac{z+\sqrt{z^2-\cos^2{\theta} }}
{\cos{\theta} } \right) \nonumber \label{gap_polar}
\end{eqnarray}
where $\theta^*=\arccos{z}$, for $z=I/\Delta <1$, and $\theta^*=0$
for $z>1$. Performing the integrations (detailed calculations will
be given elsewhere \cite{long}) we obtain the algebraic gap
equation,
\begin{eqnarray}
&& \ln\left(1/y\right)=z^3\frac{\pi}{4} ~,\hspace{1.5cm}(z<1)
\nonumber\\
&& \ln\left(\frac{1/y}{z+\sqrt{z^2-1}}\right)
=-\frac{z}{2}\sqrt{z^2-1} +\frac{z^3}{2}\arcsin[{{z}^{-1}}]~,
\nonumber\\
&& \hspace{4cm} (z>1)
\label{curve_polar}
\end{eqnarray}
where $y=\Delta/\Delta_0$ is the relative magnitude of the gap
compared to its value at $I=0$,
\begin{equation}
\Delta_0^{pol}={\rm exp}\left(-\frac{3}{\nu g}+\frac{1}{3}\right)
\approx 1.40~{\rm exp}\left(-\frac{3}{\nu g}\right).
\end{equation}
for weak coupling. There is a factor $3$ in the exponent with
anisotropic interaction instead of $1$ as in the s-wave BCS. For
small values of $z$ the solution to the gap equation is
$y=1-\frac{\pi x^3}{4}$ with $x=I/\Delta_0$. We depict the
solution of the polar phase gap equation in Fig.~1,
with the following numerical values of the characteristic
points: $x_A=(4/3\pi e)^{1/3}=0.538$, $y_A=e^{-1/3}=0.717$ 
(at the point A $y'(x)\rightarrow\infty$),
$y_C=e^{-1/3}/2=0.358$ (at the point C $y(x)=0$).

\begin{figure}[h]
\label{fig2}
\resizebox{.3\textwidth}{!}{\includegraphics{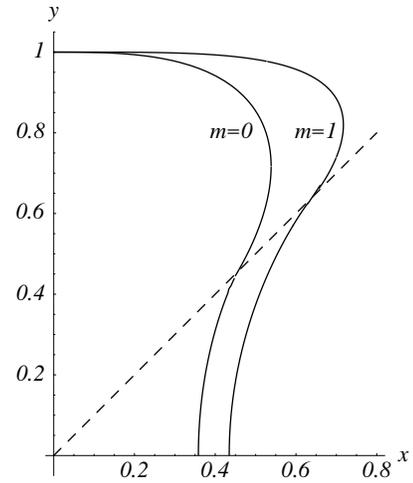}}
\caption{Solutions $y(x)$ of the gap equation in the polar, $m=0$, and
planar, $m=1$, phases.
The lower branch corresponds to the unstable state. The branches merge
at the points where $y'(x)\rightarrow\infty$, beyond which there are no 
non-zero solutions of the gap equation.
The broken line $\Delta=I$ is included to guide the eye.}
\end{figure}

{\it Planar phase:}
$\Delta_{\bf n} \sim Y_{11}({\bf n})$, $\Delta_{\bf n} \sim Y_{1-1}({\bf n})$.
We now look for a solution in the form $\Delta_{\bf n}
=\Delta\sin{({\bf n},{\bf z})e^{i\phi}}$, where $\phi$ is the
polar angle in the plane perpendicular to ${\bf z}$. The gap
equation becomes,
\begin{eqnarray}
 -\frac{2}{\nu g} &=& \int\limits_0^{\pi/2} d\theta
\sin^3{\theta} \ln{\left(\Delta\sin{\theta}\right)}
\nonumber\\
&&+ \int\limits_0^{\theta^*} d\theta \sin^3{\theta}
\ln\left(\frac{z+\sqrt{z^2-\sin^2{\theta} }}{ \sin{\theta} }
\right) \label{gap_planar}
\end{eqnarray}
where $\theta^*=\arcsin{z}$, for $z<1$, and $\theta_0=\pi/2$, for $z>1$. 
Performing the integration we obtain the algebraic gap equation,
\begin{eqnarray}
\ln{(1/y)}=-\frac{z^2}{4}+\frac{z}{8}(3+z^2)
\ln{\left|\frac{1+z}{1-z} \right|}+\frac{1}{2} \ln{|1-z^2|},
\label{curve_planar}
\end{eqnarray}
where again $y$ is the relative magnitude of the gap compared to
its value for a zero mismatch. For the planar phase,
\begin{equation}
\Delta_0^{pl}=\frac{1}{2}~ {\rm exp}\left(-\frac{3}{\nu
g}+\frac{5}{6}\right)\approx 1.15~{\rm exp}\left(-\frac{3}{\nu
g}\right).
\end{equation}
For small values of $z$ the solution to the gap equation has the
form $y=1-\frac{3x^4}{4}$ and $x$ is defined as before.
Note that the planar phase gap is more robust than the polar phase,
being perturbed by the fourth power instead of the third.
Solution of the gap equation for the planar phase is depicted in Fig.~1,
with the following numerical values of the characteristic
points: $x_A=0.674$, $y_A=0.787$, $z_A=x_A/y_A=0.856$;
$x_C=e^{-5/6}= 0.435$.

{\it Stability:}
The condensation energy is given by (at $T=0$),
\begin{eqnarray}
\Omega_s-\Omega_n&=&\nu\int\frac{d \o_{\bf n} }{4\pi}
\left(-\frac{|\Delta_{\bf n}|^2}{2}+I^2\right.\nonumber\\
&-&\left.I\sqrt{I^2-|\Delta_{\bf
n}|^2}~\Theta(I-|\Delta_{\bf n}|) \right).
\end{eqnarray}
Evaluating this expression for $z=I/\Delta<1$, we obtain for polar phase
\begin{equation}
\Omega_s-\Omega_n=\nu \Delta^2 \left(-\frac{1}{6}
-\frac{\pi z^3}{4}+z^2\right),
\end{equation}
which is negative for $z<0.537$,
and for planar phase
\begin{equation}
\Omega_s-\Omega_n=\nu \Delta^2 \left(-\frac{1}{3}
+\frac{z^2}{2}+\frac{z(1-z^2)}{4} \ln\frac{1+z}{1-z}\right),
\end{equation}
which is negative for $z<0.623$.
For our specific model Hamiltonian, at weak coupling, the planar phase
is more stable.
For $I>\Delta$ the condensation energy is always positive,
indicating that the lower branches are unstable.

Following the standard methods in the theory of superconductivity
\cite{AGD} we calculate the super-currents in our system
under the influence of homogeneous in space vector potential ${\bf
A}$.  The super-current is anisotropic, $j_i=\frac{e^2
N}{m}\kappa_{ik}A_k$ with the components given by
($\kappa_{xx}=\kappa_{yy}$),
\begin{equation}
\left[\begin{array}{l} \kappa_{zz}\\ \kappa_{xx}
\end{array}\right] = 1- \frac{3I}{2} \int\ \frac{d \o_{\bf n} }{4\pi}
\left[\begin{array}{l}\cos^2{\theta} \\
\sin^2{\theta}
\end{array} \right] \frac{\Theta(I-|\Delta_{\bf n}|)}{\sqrt{I^2-|\Delta_{\bf
n}|^2}}.
\end{equation}
For the polar phase, assuming $z>1$ we find,
\begin{equation}
\left[ \begin{array}{l} \kappa_{zz}\\ \kappa_{xx}
\end{array}\right] =\begin{array}{l}
1-3z^3/4\pi, \\
1-3\pi z/4 +3\pi z^3/8,
\end{array}
\end{equation}
The coefficient $\kappa_{xx}$ becomes negative at $z\geq 0.480$
($\kappa_{zz}$ at higher values of $z\geq 0.752$) indicating an
instability with respect to a transition into some inhomogeneous
state (probably similar to a LOFF state).
For the planar phase,
\begin{eqnarray}
\left[ \begin{array}{l} \kappa_{zz}\\ \kappa_{xx}
\end{array}\right] =
1\mp \frac{3z^2}{4} -\frac{3z}{8}(1\mp z^2)\ln\left(
\frac{1+z}{1-z}\right).
\end{eqnarray}
The coefficients $\kappa_{xx}$ and $\kappa_{zz}$ remain always positive 
for the whole range of
$z<0.623$ where the gap equation (\ref{curve_planar}) has stable
solutions.
Thus, we find that the planar phase has lower energy and higher
density of Cooper pairs than the polar phase and is therefore more
stable.

{\it Specific Heat:} The important manifestation of the BCS states
with gapless excitations  is the appearance of the term linear in
temperature in the specific heat, which is characteristic for a
normal Fermi liquid. 
The specific heat is
given by
\begin{equation}
\label{spec_heat}
 C=\sum_{{\bf p}}\left(E_{\bf p}^{+}\frac{\partial
n(E_{\bf p}^+)}{\partial T} + E_{\bf p}^{-}\frac{\partial
n(E_{\bf p}^-)}{\partial T}\right),
\end{equation}
where $E^{\pm}_{\bf p}=\pm\sqrt{\xi^2_p+\Delta_{\bf n}^2}+I$.
At low temperatures $T\ll I$ the first term in Eq.~(\ref{spec_heat})
gives an exponentially small contribution.
The second term, with $E^{-}$, in Eq.~(\ref{spec_heat}) is,
\begin{equation}
C=\frac{\nu}{4T^2}\int\limits_{-\infty}^\infty d\xi
\int\frac{d \o_{\bf n} }{4\pi}
\frac{\left(\sqrt{\xi^2+|\Delta_{\bf n}|^2}-I
\right)^2}{\cosh^2{\left[\frac{\sqrt{\xi^2+|\Delta_{\bf n}|^2}-I}{2T}\right]}}.
\end{equation}
Performing the integration,
we calculate
the contribution of the gapless modes to the specific heat 
at  $T\ll I$ to be
\begin{equation}
C=\nu T\frac{\pi}{6} \left\{
\begin{array}{ll}\pi z,& \text{polar phase},
\\ 4z^2,& \text{planar phase}.
\end{array} \right.
\end{equation}
As expected, the ``normal'' contribution to the specific heat, is
proportional to the area occupied by the gapless modes, i.e. the
$I/\Delta$ strip around the equator for the polar phase and the
$I^2/\Delta^2$ islands around the poles for the planar phase.

{\it Conclusion and Comments:}
We have presented substantial evidence that our simple model supports the planar phase gapless superfluidity
in the ground state.  For  $I\ll \Delta$ the gapless modes contribute
high powers in terms of mismatch, $\sim I^4$ for the solution
and $\sim I^2$ for the heat capacity, i.e. they represent small perturbations.   The residual continuous
symmetry of this state, and its favorable energy relative to plausible competitors (normal state, polar phase)
suggest that it is a true ground state in this model. The planar phase is symmetric under simultaneous axial rotation 
and gauge (i.e., phase) transformation.
Also, we obtain a positive density of superconducting electrons,
suggesting that inhomogeneous LOFF phases are disfavored at small $I$.

In some respects
the same qualitative behavior we find here in the p-wave resembles what arose in s-wave \cite{FGLW}.
Namely, isotropic s-wave superconductivity has two branches of solution:
the upper BCS which is stable and -- for simple interactions -- fully gapped, and the lower branch which has gapless modes
but is unstable. The striking difference is that in p-wave the upper branch retains stability while developing 
a full two-dimensional
fermi surface of gapless modes.  Thus the anisotropic p-wave breached pair phase, 
with coexisting superfluid and normal components,
is stable already for
a wide range of parameters at weak coupling using the simplest (momentum-independent) interaction.  This
bodes well for its future experimental realization.

In our model, which has no explicit spin degree of freedom,
gapless modes occur for either choice of order parameter with residual continuous symmetry.
By contrast, for $^{3}$He in the B phase the p-wave spin-triplet order parameter is a $2\times 2$ spin matrix, containing both
polar and planar phases components, there are no zeros in the quasiparticle energies, and the phenomenology
broadly resembles that of a conventional s-wave state \cite{BW}; in the A phase (which arises only at  $T\neq 0$ \cite{Leggett})
the separate up and down spin components pair with themselves, 
in an orbital P-wave, and no possibility of a mismatch arises.

Experimentally, the microscopic nature of the pairing state can be revealed most directly by probing the momentum distribution of the fermions, 
including angular dependence. Time of flight images, obtained when trapped atoms are released from the trap
and propagate freely, reflect this distribution.  

It is possible that the emergent fermi gas of gapless excitations develops, 
as a result of residual interactions, secondary condensations.
Also, one may consider analogous possibilities for particle-hole, as opposed to particle-particle, pairing.
In that context, deviations from nesting play
the role that fermi surface mismatch plays in the particle-particle case. We are actively investigating these issues.

The authors thank E. Demler, M. Forbes, O. Jahn, R. Jaffe, B. Halperin, G. Nardulli, A. Scardicchio,
O. Schroeder, A. Shytov, I. Shovkovy, D. Son, V. Liu for useful discussions.
This work is
supported in part by funds provided by the U. S. Department of Energy
(D.O.E.) under cooperative research agreement DF-FC02-94ER40818, and by NSF grant DMR-02-33773.

\end{document}